\documentclass[12pt]{iopart}
\usepackage{iopams,epsfig}  

\newcommand{\be}{\begin{equation}}
\newcommand{\ee}{\end{equation}}
\newcommand{\bd}{\begin{displaymath}}
\newcommand{\ed}{\end{displaymath}}
\newcommand{\ba}{\begin{array}}
\newcommand{\ea}{\end{array}}
\newcommand{\bt}{\begin{tabular}}
\newcommand{\et}{\end{tabular}}

\newcommand{\bea}{\begin{eqnarray}}
\newcommand{\eea}{\end{eqnarray}}
\newcommand{\bean}{\begin{eqnarray*}}
\newcommand{\eean}{\end{eqnarray*}}
\newcommand{\non}{\nonumber}

\newcommand{\Z}{\mathbb{Z}}

\newcommand{\R}{\mathbb{R}}

\begin{document}
\title{U-duality (sub-)groups and their topology}

\author{Arjan Keurentjes\dag\ }

\address{\dag\ Theoretische Natuurkunde (TENA), Vrije Universiteit
  Brussel, Pleinlaan 2, B-1050 Brussels, Belgium}

\begin{abstract}
We discuss some consequences of the fact that symmetry groups
appearing in compactified (super-)gravity may be non-simply
connected. The possibility to add fermions to a theory results in a
simple criterion to decide whether a 3-dimensional
coset sigma model can be interpreted as a dimensional reduction of a
higher dimensional theory. Similar criteria exist for higher
dimensional sigma models, though less decisive. Careful examination of
the topology of symmetry groups rules out certain proposals for
M-theory symmetries, which are not ruled out at the level of the
algebra's. We conclude with an observation on the relation between the
``generalized holonomy'' proposal, and the actual symmetry groups
resulting from $E_{10}$ and $E_{11}$ conjectures. 
\end{abstract}

\ead{arjan@tena4.vub.ac.be}


%\maketitle

\section{Introduction}
Since the construction of supergravities, and the
discovery of the Cremmer-Julia groups of compactified 11 dimensional
supergravities \cite{Cremmer:1978km, Cremmer:1979up} it has been clear
that Lie groups and algebra's play an important role in this field.

In most treatments however, the attention is confined to Lie
algebra's, and the global properties of the groups they generate are
neglected. However, a study of these global properties may lead to
useful information on the theory. In \cite{Keurentjes:2003yu} some
tools for the study of the topology of subgroups were given. Also two
applications were discussed: A criterion which contains information on
whether a theory is a dimensional reduction of a higher dimensional
one; and, a critical examination (and unfortunately, falsification) of
some proposals for symmetry groups of the yet elusive M-theory.

\section{Topology of groups}

We will not give a full discussion on the topology of groups here (the
reader is referred to \cite{Keurentjes:2003yu} and numerous textbooks on
Lie groups), but only make a few remarks.

Every simple, compact Lie group $G$ has a simply connected cover
$\widetilde{G}$. A fundamental theorem in Lie group theory states that the
group $G$ is isomorphic to $\widetilde{G}/Z$, where $Z$ is a subgroup of
the center of $\widetilde{G}$. The groups $G$ and
$\widetilde{G}$ have isomorphic Lie algebra's. Nevertheless, the effect of
the center of $\widetilde{G}$ can be seen in representation theory.

An example that is well-known to the physicist is $SU(2)$. This group
has a $\Z_2$ center, hence there are, up to isomorphism 2 different
groups with Lie algebra $su(2)$, namely $SU(2)$ and $SO(3) \cong
SU(2)/\Z_2$. There is exactly one irreducible representation (irrep)
of $SU(2)$ of dimension $n$. If $n$ is even, then this is an irrep of
$SU(2)$, but not of $SO(3)$; the mapping of $SO(3)$ to an even
dimensional irrep of $SU(2)$ is one-to-two and therefore not a homomorphism.

A similar relation is true for other compact Lie-groups: an irrep of
$\widetilde{G}$ may not give an irrep of $G$. Unlike elsewhere in the
physics literature, we will be precise in this paper; when we mention
a group $G$, it is implied that all irreps that are irreps of
$\widetilde{G}$ but not of $G$ are absent. As an example, when we say that
a symmetry group is $SO(3)$, it means that only odd-dimensional irreps
are present.

An important fact is that, even if a group is simply connected, it may
nevertheless have non-simply connected subgroups (the $SO(3)$ subgroup
of $SU(3)$, obtained by restricting to \emph{real} $SU(3)$ matrices is
a simple example). The existence of such subgroups can lead to
interesting physical effects \cite{Keurentjes99}, and is actually the
crucial ingredient in our discussion below.

\section{Fermions and oxidation}

Consider a 3 dimensional sigma model on a coset $G/H$, coupled to
gravity. An interesting question is whether this can be interpreted as
the effective theory of the toroidal compactification of a higher
dimensional theory. The answer to this question depends on the coset
$G/H$, but is often affirmative \cite{Julia:1980gr, Julia:1982gx,
  Breitenlohner:1987dg, Cremmer:1999du, Keurentjes:2002xc}. In
\cite{Keurentjes:2002xc} we showed that the possibility for oxidation
(reconstruction of the higher dimensional theory) can be deduced
from properties of $G$, and that all possible higher dimensional
theories are encoded in the geometry of the root lattice of $G$. Here
instead, we will demonstrate that also $H$ gives an immediate
criterion about the possibility of oxidation.

Consider the possibility of adding fermions to the theory. The
reduction of General Relativity from $d$ to 3 dimensions gives rise to
a 3-dimensional sigma model on $SL(d-2,\R)/SO(d-2)$
\cite{Cremmer:1999du}. The $SO(d-2)$ group appearing here can be
thought of as the remnant of the helicity group in $d$ dimensions
\cite{Julia:1980gr}. We stress that the subgroup of $SL(d-2,\R)$ is
indeed $SO(d-2)$. Now $\pi_1(SO(d-2)) = \Z_2$ (for $d > 4$), a well
known fact which is of course crucially related to the existence of
fermions. Massless fermions in the higher dimensional theory transform
in representations of $Spin(d-2)$ that are not representations of
$SO(d-2)$. In the special case $d=4$ we are dealing with $SO(2)$, and
$\pi_1(SO(2))= \Z$. Representations of $SO(2)$ are labelled by a number
(spin), and it is customary to normalize this charge such that the
bosons have integer spins. Then the fermions turn out to have
half-integer spins, and again we are dealing with a double cover of
the group relevant to the bosons.

The fact that the fermions transform in a double cover of the group
remains true after dimensional reduction. But then it is crucial that
the group $H$, appearing in the coset $G/H$ must have a topology that
is compatible with that of $SO(d-2)$. That is, the $2 \pi$ rotation
that leaves bosons invariant, but multiplies fermions with a sign,
must be represented on $H$. In mathematical language
\be
\pi_1(H) \supset \pi_1(SO(d-2),
\ee  
or, more precisely,
\be \label{neces}
\pi_1(H) \supset \Z \  (\textrm{for }d=4), \qquad \pi_1(H) \supset
\Z_2 \ (\textrm{for }d>4).
\ee

This gives a \emph{necessary} criterion for the possibility to
oxidize. An analysis of the possibilities for oxidation from coset
theories on $G/H$, with \emph{simple} $G$, indicates that it is also a
\emph{sufficient} criterion \cite{Keurentjes:2003yu}! Hence we have a

{\bf Theorem:} Consider a sigma model in 3 dimensions on a symmetric
space $G/H$, with $G$ a simple non-compact group and $H$ its maximal compact
subgroup, coupled to gravity. This sigma model can be oxidized to a
higher dimensional model if and only if the group $H$, as embedded in
$G$, is not simply connected. Moreover, the maximal oxidation
dimension $d$ is given by: 
\be
\ba{lll}
d=3 & \textrm{if} & \pi_1(H) = 0; \\
d=4 & \textrm{if} & \pi_1(H) = \Z; \\
d > 4 &  \textrm{if} & \pi_1(H) = \Z_2. \\
\ea
\ee

A full list of cosets can be found in \cite{Keurentjes:2003yu}. Here
we restrict to examples that are related to cosets of 3 dimensional
\emph{supergravity} theories. For sufficiently many supersymmetries,
the target space geometry of the sigma model must be a symmetric space
\cite{deWit:1992up}. A priori, the constraints of 3-dimensional
supergravity are not related to spin (which does not
exist in 3 dimensions). The theory of oxidation
\cite{Keurentjes:2002xc} and the considerations on fermions
\cite{Keurentjes:2003yu} provide the link between the analysis of
\cite{deWit:1992up}, and the analysis in higher dimensional theories
(which are restricted to have not more than one spin 2 excitation).

\begin{table}
\caption{Coset symmetries $G/H$ for 3-d supergravity theories with $N$
  supersymmetries. For $N < 9$ there is extra freedom parametrized by
  an integer $k$. The maximal oxidation dimension is denoted by
  $d$.}\label{tab1} 
\begin{indented}
\item[]\begin{tabular}{@{}cccc}
\br
$N$ & $d$ & $G$ & $H$ \\
\mr
16 & 11 & $E_{8(8)}$ & $Spin(16)/\Z_2$ \\
12 &  6 & $E_{7(-5)}$ &  $(Spin(12)\times SU(2))/\Z_2$  \\
10 &  4 &  $E_{6(-14)}$  &   $Spin(10) \times U(1)$ \\
9  &  3 &  $F_{4(-20)}$  &  $Spin(9)$  \\
8,  $k > 2$   & $\min(10,k+2)$  &  $Spin(8,k)$  &  $(Spin(8)\times
Spin(k))/\Z_2$ \\ 
8,  $k=2$  & 4 &  $Spin(8,2)$  &  $Spin(8) \times U(1)$  \\
8,  $k=1$  & 3 &  $Spin(8,1)$  &  $Spin(8)$ \\
6,  $k$    & 4 &  $SU(4,k)$  &  $S(U(4) \times U(k))$ \\
5,  $k$    & 3 &  $Sp(2,k)$  &  $Sp(2) \times Sp(k)$ \\
\br 
\end{tabular}
\end{indented}
\end{table}

Table \ref{tab1} was taken from \cite{deWit:1992up} (but note a
few corrections and the adaptation to our standards). Note: That our
criterion confirms that theories with an odd number of 3-d
supersymmetries cannot be oxidized (simply connected $H$); that
theories which can be oxidized to 4 dimensions have a single $u(1)$
factor in their $H$-algebra; and that for all theories that can be 
oxidized to higher dimensions (for which $N$ is a multiple of 4, and
which may require suitable matter content), the
fundamental group of $H$ is $\Z_2$.

The reader may have noticed that a similar reasoning can be set up for
theories in higher dimensions. Consider a $d$ dimensional theory, with
a sigma models on $G/H$. If the theory can be derived as a dimensional
reduction of a yet higher-, $d+D$ dimensional theory, then the group
$H$ has to contain the group $SO(D)$, and moreover
\be
\pi_1(H) \supset \pi_1(SO(D)).
\ee
It should be emphasized that, in contrast to the 3-dimensional case,
this is really not more than a (rather weak) necessary criterion, as
counterexamples to sufficiency are numerous (e.g. IIB supergravity in
10 dimensions, with $SL(2,\R)/SO(2)$).

\section{Generalized holonomy and symmetries of maximal supergravities}

In \cite{Duff:2003ec, Hull:2003mf} a ``generalized holonomy'' proposal
was put forward. The reasoning behind this proposal is roughly as
follows.

There exist formulations of dimensionally reduced maximal supergravity
with local symmetry $Spin(1,d-1) \times \widetilde{H}_d$. Here
$Spin(1,d-1)$ is obviously the local Lorentz-group. The second factor
represents the double cover of a maximal compact subgroup of a
Cremmer-Julia group \cite{Cremmer:1979up} (see table \ref{tab2} for a
list of these). The existence of these ``hidden'' symmetries prompts
the question whether these are a consequence of compactification, or
already present in some form in the higher dimensional theory. An
answer to this question was given in \cite{deWit:1985iy}, where
formulations of \emph{11 dimensional supergravity} with local
$Spin(1,d-1) \times \widetilde{H}_d$ invariance were constructed.

These symmetries are local, and presumably also symmetries of the
proposed non-perturbative extension of 11-d supergravity,
M-theory. Upon compactification, such symmetries are broken by
boundary conditions; more accurately, there is non-trivial
\emph{holonomy} in the group $Spin(1,d-1) \times \widetilde{H}_d$,
such that it is no longer a manifest symmetry of the lower dimensional
theory.

The groups $Spin(1,d-1) \times \widetilde{H}_d$ refer to a specific
factorization of the background geometry, into a $d$-dimensional part,
containing the time-like direction, and an $(11-d)$ dimensional
part. For a full description, one wants to know $\widetilde{H}_d$ for
all values of $d$, specifically for $d = 0$. For $d \geq 3$, these
groups are known from the Cremmer-Julia analysis
\cite{Cremmer:1979up}, and \cite{deWit:1985iy}. For $d=2,1$ the groups
$Spin(16) \times Spin(16)$ and $Spin(32)$ were proposed by
\cite{Duff:2003ec}, for $d=0$ a proposal is $SL(32,\R)$
\cite{Hull:2003mf}. We will here re-examine these proposals more
carefully.

\begin{table}
\caption{For $d \geq 3$: Cremmer-Julia groups $G_d$; their compact
  subgroups $H_d$. For $d <3$: Candidate ``generalized
  holonomy''-groups in lower dimensions} \label{tab2}
\begin{indented}
\item[] \begin{tabular}{ccc}
\br
d & $G_d$ & $H_d$ \\
\mr
11 & $\{e\}$ & $\{e\}$ \\
10 & $\R$, $SL(2,\R)$  & $\{e\}, SO(2)$  \\
9  & $SL(2,\R) \times \R$ & $SO(2)$ \\
8  & $SL(3,\R) \times SL(2,\R)$ & $SO(3) \times SO(2)$ \\
7  & $SL(5,\R)$ & $SO(5)$ \\
6  & $Spin(5,5)$ & $(Sp(2) \times Sp(2))/\Z_2$ \\
5  & $E_{6(6)}$ & $Sp(4)/\Z_2$ \\
4  & $E_{7(7)}$ & $SU(8)/\Z_2$ \\
3  & $E_{8(8)}$ & $Spin(16)/\Z_2$ \\
\mr
2  &            & $Spin(16) \times Spin(16)$ \\
1  &            & $Spin(32)$               \\
0  &            & $SL(32,\R)$            \\
\br
\end{tabular}
\end{indented}
\end{table}
In table \ref{tab2} we have collected the Cremmer-Julia groups, and
their actual compact subgroups. Our table differs from many others in
the literature because we have been careful to mention the compact
groups $H_d$ with their correct topologies. The reader will notice
that for $d < 8$, all $H_d$ are simple, and \emph{two-fold connected}.

The two-fold connectedness is related, as before, to fermionic
representations. The bosons in the supergravity theories transform in
irreps of $H_d$. The fermions however, transform in irreps of the
double cover $\widetilde{H}_d$ that are not irreps of $H_d$. An
important fact to keep in mind is that, since $\widetilde{H}_d$ is not
a subgroup of $G$, $G$ can represent at most symmetries from the
bosonic sector of the theory. Not only do the fermions not transform
in irreps of $G$, there does not even exist a $G$ representation that
has the fermionic irreps in its $H_d$ decomposition.

Nevertheless, since there are fermions present in the theory, the full
symmetry of the theory contains $\widetilde{H}_d$. The example that
will be important to us is 3-dimensional maximal supergravity
\cite{Marcus:1983hb}. The compact subgroup of $E_{8(8)}$ is
$Spin(16)/\Z_2$. The scalars in the theory are in the
$\mathbf{128_s}$, which is an irrep of $Spin(16)/\Z_2$. The
(non-dynamical) gravitini are in the $\mathbf{16}$, whereas the
remaining fermions are in the $\mathbf{128_c}$ (the \emph{other} spin
irrep of $Spin(16)$). Neither of the latter 2 irreps is an irrep of
$Spin(16)/\Z_2$, and hence the full symmetry of the theory is
$\widetilde{H}_3 = Spin(16)$.

We now turn to the proposed generalized holonomy groups for $d <
3$. By restriction to a smaller number of ``internal''dimensions, we expect
\be
\widetilde{H}_d \supset \widetilde{H}_{d+1}.
\ee 
Hence, we expect
\be \label{embed}
SL(32,\R) \supset Spin(32) \supset Spin(16) \times Spin(16) \supset
Spin(16).
\ee
The crucial point however is that equation \ref{embed} is \emph{false}!
The actual subgroup of $SL(32,\R)$ with $so(32)$ algebra is $SO(32)$,
and not $Spin(32)$: No spin irreps of $Spin(32)$ can appear in the
decomposition from $SL(32,\R)$ irreps. In turn, the subsequent
subgroup of $SO(32)$ is $SO(16) \times SO(16)$, and this subgroup has
only $SO(16)$ subgroups. It is impossible to obtain the irreps
$\mathbf{128_s}$ and $\mathbf{128_c}$ of the scalars and fermions in
3-d supergravity, from any $SL(32,\R)$ irrep.

Some more thought reveals that $SL(32,\R)$ has no $Spin(16)$ subgroups
whatsoever \cite{Keurentjes:2003yu} also if we are willing to give up
the chain in equation \ref{embed}. The inevitable conclusion is then
that 3-d supergravity has symmetries not contained in $SL(32,\R)$
(related to the center of $Spin(16)$) which can therefore not be a
symmetry group in the sense proposed in \cite{Hull:2003mf}.

A similar, but slightly more subtle reasoning applies to
$Spin(32)$. The subgroup of $Spin(32)$ with $so(16) \oplus so(16)$
algebra is $(Spin(16) \times Spin(16))/\Z_2$. This group has various
subgroups with $so(16)$ algebra. There are essentially two
options, an embedding in one of the two factors, or the diagonal
one. The proposal in \cite{Duff:2003ec} claims that the 2-d gravitini
should be in the $\mathbf{(16,1) \oplus (1,16)}$. Then, to get the
proper $\mathbf{16}$ for the 3-d gravitini, we should select the
diagonal embedding. But the diagonal subgroup in $(Spin(16) \times
Spin(16))/\Z_2$ is $SO(16)$, and we again find a
contradiction. Alternatively, a non-diagonal embedding leads to 16
singlets that do not fit in the 3-d theory, that furthermore would
transform in an (unobserved) extra $Spin(16)$ factor. As it seems
impossible to make sense out of this, we discard $Spin(32)$ as
candidate group for $\widetilde{H}_1$.  

We cannot rule out $Spin(16) \times Spin(16)$ on the basis of these
arguments, but it is clear that there is reason to distrust this group
too. Indeed, a careful analysis of 2-d maximal supergravity, as
performed in \cite{Nicolai:1998gi} does not indicate this symmetry.

Though the abstract ``generalized holonomy'' proposal is attractive,
the precise symmetry groups proposed seem to be ruled out. It is not
easy to find alternative, finite-dimensional candidates. Instead,
conjectures on $E_{10}$ and $E_{11}$ symmetries in maximal
supergravity seem to indicate that the local symmetry groups should be
infinite dimensional. The next section  demonstrates a link between
these infinite dimensional groups, and the discarded ``generalized
holonomy groups''.

\section{Generalized holonomy and infinite dimensional groups}

In spite of the fact that the proposal that these groups are symmetry
groups for M-theory turns out to be untenable, there is a very simple
relation between the groups proposed, and the $E_{n(n)}$ groups of
maximal supergravities, also for $n=10,11$ ($d=1,0$), where the
symmetry groups have a conjectural status.
\begin{figure}[ht]
\begin{center}
\includegraphics[width=17cm]{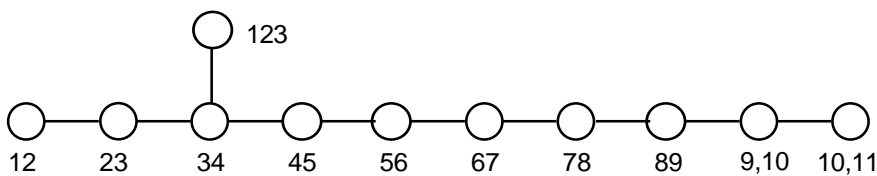}
\caption{The Dynkin diagram of $E_{11}$.}\label{e11fig}
\end{center}
\end{figure}

In figure \ref{e11fig} we have depicted the Dynkin diagram of $E_{11}$,
but the discussion below extends to any $E_n$ group with $n <11$ by suitably
truncating the Dynkin diagram (it also extends to $E_n$ with $n > 11$,
but this is without obvious application to supergravity). We have
labelled all nodes with a set of integers, the nodes along the
horizontal line with a pair of integers, the branch with a triplet of integers.

The ``generalized holonomy groups'' from \cite{Duff:2003ec, Hull:2003mf}
are obtained as follows. To obtain the relevant group for $d$
dimensions, we omit all nodes that have numbers larger
than $d$ in their index set from the diagram. For each of the
remaining nodes, we form the Clifford algebra element $\Gamma^{S_i}$,
where $S_i$ is the index set coming with node $i$. As usual
$\Gamma^{S_i}$ is a product of gamma matrices, fully anti-symmetrized
in the indices. Next we form the algebra of consecutive
commutators of the $\Gamma^{S_i}$. This algebra in turn generates a
Lie group $H_{\Gamma}$, which is the ``generalized holonomy group''
mentioned in \cite{Duff:2003ec,Hull:2003mf}. The groups for time-like,
space-like and null reduction follow from including in the set of generating
gamma matrices an element that squares to $-1,1,0$, respectively.

The reason for presenting the ``generalized holonomy groups'' like
this, is that the above construction has clear parallels with the abstract
construction of Lie algebra's \cite{Kac:gs}. Also there the Lie
algebra is defined by forming consecutive commutators of ladder operators
$e_i$. Another set of generators consists of consecutive commutators
of the conjugate ladder operators $f_i$. Together with the Cartan
generators $h_i$, the $e_i$ and $f_i$ generate the full algebra.

The factor group $H$ appearing in $E_{n(n)}/H$ is generated by
elements of the form $e_i - \epsilon_i f_i$, and their commutators. The
$\epsilon_i=1,-1,0$ is included to allow for other than spacelike
reductions (if a null or timelike direction is present we choose this to be the
$d$-direction; $\epsilon_i=0(-1)$ for a node $i$ if the index
set associated to it contains a null (timelike) direction; otherwise
$\epsilon_i=1$.). 

For \emph{finite-dimensional} $E_{n(n)}$ we have \emph{exactly} 
\be
H \cong H_{\Gamma}.
\ee
There is a one-to-one relationship between $e_i - \epsilon_i f_i$ and
$\Gamma^{S_i}$. Furthermore $\epsilon_i = - (\Gamma^{S_i})^2$ (where
the left hand side includes the identity on the spinor algebra).

This is however not so for $n > 8$. The group $H_{\Gamma}$ is always a
\emph{finite-dimensional} group, while for $n > 8$ the group $H$ is
clearly infinite. Nevertheless, there is still the relation between
the generating elements of $H$, and the $\Gamma^{S_i}$. If we, as in
\cite{Duff:2003ec}, denote symbolically by $\Gamma^{(n)}$ the Clifford algebra
elements obtained by multiplying $n$ gamma-matrices and
anti-symmetrization in the indices, then we have:
\bea
\left[\Gamma^{(3)}, \Gamma^{(3)} \right] = \Gamma^{(2)} + \Gamma^{(6)};
& \qquad & 
\left[\Gamma^{(3)}, \Gamma^{(6)} \right] = \Gamma^{(3)} + \Gamma^{(7)}; \\
\left[\Gamma^{(3)}, \Gamma^{(7)} \right] = \Gamma^{(6)} +
\Gamma^{(10)}; & \qquad & 
\left[\Gamma^{(3)}, \Gamma^{(10)} \right] = \Gamma^{(7)} +
\Gamma^{(11)}; \ldots  \non
\eea
In terms of the level expansion by the ``exceptional root'' (as
proposed in \cite{Damour:2002cu}),
$\Gamma^{123}$ corresponds to a linear combination of a ladder
generator $e_k$ at level 1 and one $f_k$ at level $-1$. Similarly, all
$\Gamma^{(3)}$ correspond to level $\pm1$ generators,
$\Gamma^{(6)}$ to level $\pm2$ generators, $\Gamma^{(7)}$ to level
$\pm3$ generators, and so on. It is now easy to see that the algebra
generated by the $\Gamma^{S_i}$ corresponds to the algebra $H$
\emph{truncated} beyond level $\pm (2k-1)$ for $n < (4k+2)$, and
beyond level $\pm 2k$ for $n < (4k+3)$. Furthermore,
$SO(n)$-representations other than completely antisymmetric tensors are
\emph{excluded} (because the Clifford property of
gamma-matrices contracts all symmetrized indices). 

For $n < 9$ the level truncation imposes no restriction and all irreps
are antisymmetric tensors (see \cite{Keurentjes:2002xc}). For $n=11$,
one precisely finds the generators mentioned in \cite{West:2003fc}
(with additional ones that are actually needed to complete the group to
$SL(32,\R)$). It should be stressed however, that $sl(32,\R)$ is \emph{not} a
sub-algebra of $e_{11}$ (nor is $so(32)$ a subalgebra of $e_{10}$, or
$so(16) \oplus so(16)$ a sub-algebra of $e_9$): The truncation implied
by the $\Gamma^{S_i}$ is an \emph{illegal} procedure when selecting
sub-algebra's! 

Whether the relation exhibited here has any other
profound consequences, we leave for future research and speculation.

\ack
I owe many thanks to Hermann Nicolai and
Chris Hull for discussions and extensive correspondence, before and
after publication of \cite{Keurentjes:2003yu}. The author is
supported in part by the ``FWO-Vlaanderen'' through project G.0034.02,
in part by the Federal office for Scientific, Technical and Cultural
Affairs through the Interuniversity Attraction Pole P5/27 and in part
by the European Commision RTN programme HPRN-CT-2000-00131, in which
the author is associated to the University of Leuven.

\end{document}